# Measurement theory of a density profile of small colloids around a large colloid: Conversion of force between two-large spheres into pressure on the surface element

Ken-ichi Amano, Kota Hashimoto, and Ryosuke Sawazumi

*Department of Energy and Hydrocarbon Chemistry, Graduate School of Engineering, Kyoto University, Kyoto 615-8510, Japan.*

Author to whom correspondence should be addressed: Ken-ichi Amano.
Electric mail: amano.kenichi.8s@kyoto-u.ac.jp

**ABSTRACT**

We suggest a transform theory for calculating a density distribution of small colloids around a large colloid from a force curve between the two-large colloids. The main idea (calculation process) is that the force curve between the two-large colloids is converted into the pressure on the surface element of the large colloid. This conversion is different from the celebrated Derjaguin approximation. A numerical matrix calculation is performed in the conversion to calculate it more precisely. Subsequently, the pressure on the surface element is transformed into the density distribution of the small colloids around the large colloid by using a transform theory for surface force apparatus proposed by Amano. In this letter, the process of the transformation is explained and a prototype result of the transformation is shown.

**MAIN TEXT**

We recently proposed a measurement theory that transforms a force curve between two-large colloids into the density distribution of the small colloids around it [1], the idea of which is based on both the superposition of the radial density distributions. The force curve is an input for the transform, which can be obtained by using laser tweezers (LT) [2-8], surface force apparatus (SFA) [9], or colloid-probe atomic force microscopy (colloid-probe AFM) [11-14]. In this letter, we suggest a transform theory with a different route. The calculation process is that the force curve between the two-large colloids is converted into the pressure on the surface element of the large colloid. This conversion is different from the celebrated Derjaguin approximation [13-15]. A numerical matrix calculation is performed in the conversion to calculate the pressure on the surface element more precisely. Subsequently, the pressure on the surface element is transformed into the density distribution of the small colloids around the large colloid by using a transform theory for SFA proposed by Amano [16,17].

Derjaguin approximation has been applied in many researches [13-15,18-21] due to its universality and validity. However, the applicability of it is limited to large particles and it is restricted to very short surface-surface separation. Then, Bhattacharjee and Elimelech proposed a more precise method named surface element integration (SEI) [22,23]. SEI can convert interaction energy per unit area between two parallel flat surfaces into the energy of an entire surface of a particle with the flat surface through the integration. Certainly SEI is a very beneficial method, however, it cannot convert interaction between two spheres or between a flat surface and a sphere into an interaction between the two flat surfaces. (It seems that the difficulty lies in the mathematical or numerical analysis of the integral equation of SEI.) In the present letter, however, we need to convert the force acting on the large colloid into the pressure on the surface element. Thus, we proposed a method named FPSE conversion which can perform such the conversion. Here, FPSE means 'force to pressure on surface element', and the conversion is constructed in a granular system and uses matrix operation.

In what follows, FPSE conversion is explained, and after it one example of the transformation from the pressure on the surface element into the density distribution of the small colloids around the large colloid is explained [16]. The system configuration is shown in Fig. 1, which is constructed in the granular system. There are many small colloids with number density $\rho_0$ and large colloids 1 and 2. The solution is inert background. A space which the center of the small colloid cannot enter is excluded volume of the large colloid, and $r$ is the radius of the excluded volume. $r_S$ and $r_B$ are radii of the small and large colloids, respectively. (When $r_B \gg r_S$, $r \approx r_B$.) $\theta$ denotes the radian from the upward-vertical line originating from the large colloid 2. The separation between the centers of the large colloids 1 and 2 is represented as $s$. The length of the horizontal line between the excluded surfaces is represented as $l$. If the force between the large colloids 1 and 2 ($f$) is expressed by a summation of forces between face-to-face surface elements (see closed circles in Fig. 1), the force can be written as

$$f(s) = \sum_l P(l) A_{2z}(l; s), \tag{1}$$

where $P$ is pressure and $A_{2z}$ is an efficient area of the surface element of the large colloid 2, which is normal to $z$-axis. In the present case, Eq. (1) can be rewritten as

$$f(s) = 2\pi r^2 \int_0^{\frac{\pi}{2}} P(l) \sin\theta \cos\theta d\theta. \tag{2}$$

Here, we would to mention that 'pressure from left' pluses 'that from right sides' is $P$ (see Fig. 1). Moreover, we mention that meaning of the pressure $P$ is both 'that between the wall elements' and 'that on the surface element'. By the way, $l$ can be expressed as

$$l = s - 2r\sin\theta, \tag{3}$$

and hence following two expressions are obtained:

$$\cos\theta d\theta = -(1/2r)dl, \tag{4}$$

$$\sin\theta = (s - l)/(2r). \tag{5}$$

Thus, Eq. (2) is rewritten as

$$2f(s)/\pi = \int_{s-2r}^{s} P(l)(s-l)dl. \tag{6}$$

It can be seen that Eq. (6) is in the form of a matrix calculation as follows:

$$\boldsymbol{F}^{*} = \boldsymbol{H}\boldsymbol{P}, \tag{7}$$

where $\boldsymbol{F}^{*}$ corresponds to left-hand side of Eq. (6). $\boldsymbol{P}$ and $\boldsymbol{H}$ correspond to $P(l)$ and the other parts, respectively. $\boldsymbol{H}$ is a square matrix whose variables are $l$ and $s$, however, its lower right area is composed of a square unit matrix. $\boldsymbol{P}$ is numerically calculated by using, for example, the inverse matrix of $\boldsymbol{H}$. Consequently, $P(l)$ is obtained and FPSE conversion is finished.

Density distribution of the small colloids around the large colloid can be calculated by using the transform theories for two-flat surfaces [16,17]. If the system can be approximated by a rigid one, the density distribution is obtained as follows [16]:

$$g_{\text{C}} = \frac{1 + \sqrt{1 + 4P(0)/(k_{\text{B}}T\rho_{0})}}{2}, \tag{8}$$

$$g(r+l) = \frac{P(l)}{k_{\mathrm{B}}T\rho_0 g_{\mathrm{C}}} + 1, \qquad (9)$$

where $g_{\mathrm{C}}$ and $g$ are normalized number density of the small colloids at the contact and around the large colloid, respectively. $k_{\mathrm{B}}$ and $T$ are the Boltzmann constant and absolute temperature, respectively. (More specifically, the transform theory [16] requires the two-body potential between the large and small colloids is rigid one, but the others are not restricted to rigid potentials.) When the potential between the flat surface and the small colloid is approximated as soft potential with rigid wall, a theory written in [17] is recommended to use (the explanation is skipped here).

The transform theory above can obtain not only the radial density distribution of the small colloids around the large colloid, but also the three-dimensional (3D) density distribution of the small colloids around the two-large colloids. The latter is calculated by substituting the two distributions around the large colloids 1 and 2 into Kirkwood superposition approximation [1,16,17,24-26].

We verified the transform theory explained here by using a computer. At first, the normalized density distribution of the small colloids around the large colloid ($g_{\mathrm{B}}$) is prepared by using one-dimensional Ornstein-Zernike equation coupled with hypernetted-chain closure (1D-OZ-HNC), previously. Here, $g_{\mathrm{B}}$ is a benchmark for this verification test. Next, the input datum (force curve between the two large colloids) is calculated also by using 1D-OZ-HNC. (The system is modeled as rigid and the solvent here is inert background.) The input datum $f$ is converted to the pressure $P$, and $P$ is transformed into the density distribution $g$ of the output. We have found that the output $g$ is similar to $g_{\mathrm{B}}$ (not shown). Furthermore, it is found that the $g$ is *very* similar to (almost same as) the normalized density distribution calculated through the process in [1]. The theoretical relation between them will be explained in another paper.

In summary, we have proposed and explained the transform theory. At first, the force curve is converted into the pressure on the surface element by using FPSE conversion. Secondly, the pressure is transformed into the density distribution of the small colloids around the large colloid. We have briefly confirmed the validity of the transform theory, and concluded that the theory is valid. In the near future, we will conduct detailed verification of the theory and find the applicable range of it. In addition, we will present a similar transform theory for colloid probe AFM by applying both FPSE conversion and the transform theory for SFA [16,17]. We remark that the theory proposed here has potential to become a fundamental theory for measurements of the density distributions. The density distributions include: (A) radial-density distribution of the small colloids around the large colloid; (B) 3D-density distribution of the small colloids confined by the two-large colloids; (C) radial-density distribution of solvent molecules around a spherical solute; (D) 3D-density distribution of the solvent molecules confined by the two-spherical solutes. The experimental outline is as follows. The force curve between two-large particles is obtained by using LT, SFA, AFM, fluorescent microscope (FM), or light scattering (LS). When a line optical tweezer (one of the LTs), FM, or LS is used, the pair distribution function between the two-large particles ($g_{PP}$) is obtained at first. Since $g_{PP}$ has a relation with its potential of mean force ($\varphi_{PP}$) [27,28],

$$g_{PP} = \exp[-\varphi_{PP}/(k_B T)], \tag{10}$$

$\varphi_{PP}$ can readily be obtained from $g_{PP}$. Subtracting its two-body potential ($u_{PP}$) from $\varphi_{PP}$ [8], the input force (small particles mediated force) is obtained as follows:

$$f = -\frac{d}{ds}(\varphi_{PP} - u_{PP}). \tag{11}$$

After $f$ was obtained, the transform theory proposed here or another one [1] is applied. Then, $f$ is transformed to $g$ of the radial-density distribution of the smaller particles around the larger particle. If one needs 3D density distribution of the smaller particles around the two-larger particles, Kirkwood superposition approximation [1,16,17,24-26] is reused. This is the outline of the theory for experimental uses. As a first step application, potential of mean force between polymethylmethacrylate (PMMA) particles [8] is transformed into density distribution of polystyrene (PS) particles around a PMMA particle in Fig. 2. The details are written in the figure caption.

ACKNOWLEDGEMENTS

We appreciate Tetsuo Sakka (Kyoto University), Naoya Nishi (Kyoto University), and Masahiro Kinoshita (Kyoto University) for the useful advice, discussions, and data. This work was supported by "Grant-in-Aid for Young Scientists (B) from Japan Society for the Promotion of Science (15K21100)".

**FIGURES**

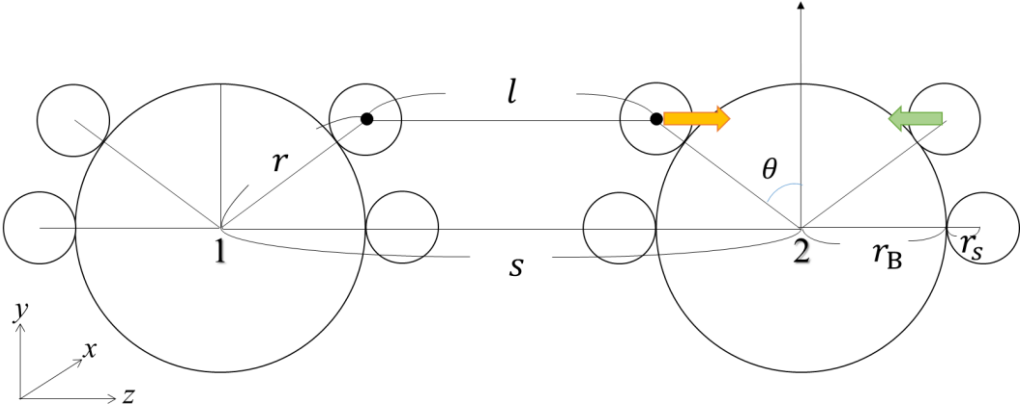

Fig. 1: The system configuration. There are many small colloids with number density $\rho_0$ and large colloids 1 and 2. The solution is inert background. A space which the center of the small colloid cannot enter is excluded volume of the large colloid.

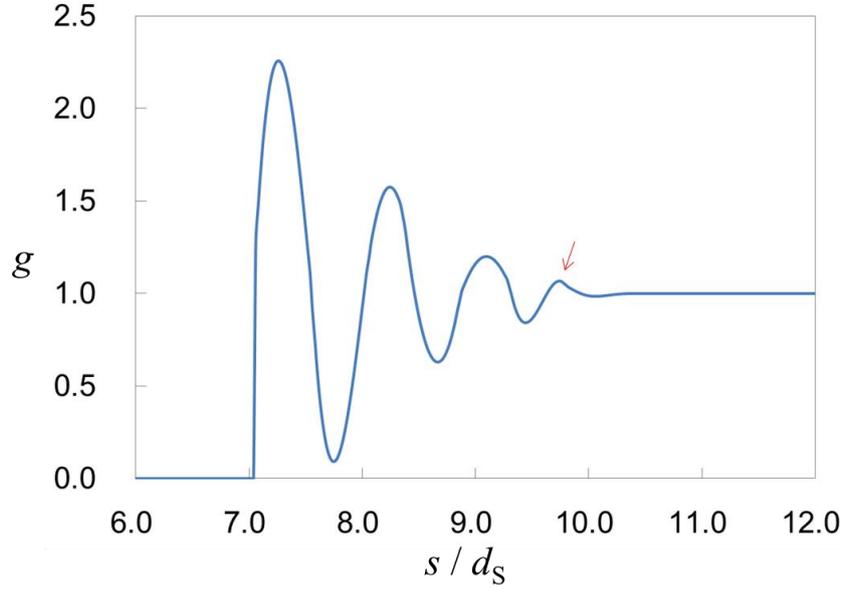

Fig. 2: Normalized density distribution of PS particles around a PMMA particle. Potential of mean force between PMMA particles in a colloidal solution measured by line optical tweezers [8] is used as an input. The colloidal solution contains many PS particles, packing fraction ($\varphi_S$) of which is 0.34. The diameters of PS and PMMA particles are 86 nm and 1131 nm, which are determined from the highest peak position of the mean force, their product diameters, and 3 nm screening length of mutual electrostatic repulsions. The calculation process is as follows. (a) Potential of mean force of its partial area is smoothly fitted by using a sextic function. (b) Four fitting curves are smoothly connected by using quantic functions, where length of the connected area is $d_S/2$ and the centers of the connected areas are 1282.4 nm, 1352.3 nm, and 1432.6 nm. (c) Tail of the fitting curve is multiplied by a soft step function $\exp(-(s/t)^{40})$, where $t$ is a converged point 1442.3 nm. (d) The fitting curve is changed to a force curve between PMMA particles. (e) The force curve is transformed into the density distribution of PS particles around a PMMA particle. Two-body potential between PMMA particles is approximated by 0 within the transformed range. If a very small peak at near 1.40 μm is absent or noise in FIG. 1 ($\varphi_S$ = 0.34) of [8], a peak at near $9.74 d_S$ (arrowed) is removed.